\documentclass{Interspeech}

\interspeechcameraready

\title{``Dyadosyncrasy'', Idiosyncrasy and Demographic Factors in Turn-Taking}

\author[affiliation={1,2,3}]{Julio Cesar}{Cavalcanti}
\author[affiliation={1}]{Gabriel}{Skantze}

\affiliation{}{KTH Royal Institute of Technology}{Sweden}
\affiliation{}{Pontifical Catholic University of São Paulo}{Brazil}
\affiliation{}{Stockholm University}{Sweden}

\email{jcca@kth.se, skantze@kth.se}

\usepackage{comment}
\usepackage{graphicx}

\begin{document}

\maketitle

\begin{abstract}

Turn-taking in dialogue follows universal constraints but also varies significantly. This study examines how demographic (sex, age, education) and individual factors shape turn-taking using a large dataset of US English conversations (Fisher). We analyze Transition Floor Offset (TFO) and find notable inter-speaker variation. Sex and age have small but significant effects -- female speakers and older individuals exhibit slightly shorter offsets -- while education shows no effect. Lighter topics correlate with shorter TFOs. However, individual differences have a greater impact, driven by a strong idiosyncratic and an even stronger ``dyadosyncratic'' component -- speakers in a dyad resemble each other more than they resemble themselves in different dyads. This suggests that the dyadic relationship and joint activity are the strongest determinants of TFO, outweighing demographic influences.

\end{abstract}

\section{Introduction}

Humans spend a substantial portion of their lives engaged in spoken interactions, seamlessly coordinating conversations through finely tuned turn-taking mechanisms. These mechanisms, central to human communication, display strikingly universal characteristics across languages -- rooted in a shared cognitive and linguistic infrastructure \cite{levinson2016, stivers2009}. Notably, the rapid transition between conversational turns -- often around 200 ms -- highlights the efficiency of human speech processing and the predictable structure of dialogue \cite{levinson2015}.

Yet, despite these universal patterns, turn-taking is far from monolithic. Cultural, developmental, and situational factors introduce considerable variation in how turn-taking unfolds across different contexts and speakers. For instance, previous studies \cite{stivers2009} demonstrated that while turn transition timings generally fall within a narrow range across ten languages, the mean gap durations vary markedly: Japanese speakers exhibit an average gap of only 7 ms, in contrast to Danish speakers, who average 489 ms.

Developmental trajectories further underscore this variability. Research indicates that children gradually acquire turn-taking skills, initially displaying longer gaps -- often between 1.5 and 2 seconds -- before refining their ability to anticipate turn completions and respond more promptly as their linguistic and social competencies mature \cite{ervin1979, casillas2016}.

Sex differences, though long recognized, have rarely been examined with contemporary quantitative methods and extensive datasets. Evidence suggests that in same-sex interactions, men tend to adhere closely to the conversational model proposed in \cite{sacks1974} -- minimizing gaps and overlaps -- while women’s conversations feature significantly more overlap \cite{ervin1979}. In a study of multi-party human–robot interactions, adult females not only dominated speaking time but also exhibited a higher degree of overlap (27\% compared to 13\% for males), an effect further amplified when two females interacted \cite{skantze2017}. Corpus studies have further confirmed that such demographic differences are part of broader ``turn-taking styles'', with duration patterns correlated to sex, age, dyadic influence \cite{grothendieck2011}, and other dimensions \cite{ward2023}.

Other factors are also known to shape turn-taking dynamics. Competitive conversational contexts tend to produce shorter gaps \cite{trimboli1984}, whereas complex tasks, unfamiliar topics \cite{bull1998}, and increased cognitive load may be associated with longer between-speaker intervals in conversation \cite{heldner2011}. Eye contact plays a role as well. Pauses and overlaps tend to exhibit distinct statistical properties in telephone versus face-to-face dialogues. In comparable conditions, pauses in face-to-face conversations can last up to four times longer than those in telephone dialogues \cite{ten2004}.
In addition, speakers tend to provide answers more rapidly than non-answer responses to questions. Among these answers, confirmations are typically produced faster than disconfirmations, and this pattern appears to hold across different languages \cite{stivers2009}. Familiarity between speakers similarly impacts conversation dynamics, as conversations between friends tend to include longer gaps \cite{templeton2023}. Personality traits further shape conversational style \cite{cuperman2009}; extroverts, for instance, overlap more frequently with their conversational partners compared to introverts \cite{yu2019}.

On a higher level, there is also the impact of the Dyadic factor on turn-taking. In fact, a considerable number of studies highlight the intricate ways in which speakers mutually influence each other’s speech patterns when engaging in conversation \cite{benus2011, levitan2011, edlund2009, kousidis2009} or even when reading together \cite{cummins2004, cummins2006}. Accommodation processes -- the dynamic adjustment whereby individuals gradually align their behaviors -- can further contribute to inter-speaker similarity \cite{benus2011}.

In conversation, speakers tend to become more like each other in different features and different ways. This phenomenon, and all other terms that may get grouped together under the word ``entrainment'', is widely believed to be crucial to the success of human interactions \cite{levitan2011}. For instance, speakers adapt the duration of gaps to those of the other participants, that is, a form of accommodation \cite{edlund2009, kousidis2009}. Speakers may exhibit convergence through proximity, synchrony, or other forms of speech entrainment, with turn exchanges often displaying heightened similarity \cite{benus2011, levitan2011}.

Collectively, these findings demonstrate that while turn-taking is governed by universal constraints, it also exhibits notable variations at both the individual and dyadic levels -- variations that have yet to be comprehensively explored and quantified. Although previous studies have significantly advanced our understanding of turn-taking, some gaps remain -- particularly attempts at quantifying the relative contribution of demographic, individual, and dyadic factors.

Our analysis adopts a corpus-based approach and focuses on the Transition Floor Offset (TFO) -- a metric that captures the temporal gap or overlap between consecutive turns, and employs mixed-effects modeling to account for both individual and conversation-level variability. We specifically ask: How do demographic differences modulate the timing of turn transitions? How variable (idiosyncratic) can it be across different speakers? What role do dyadic interactions play in amplifying or reducing inter-speaker differences?

The answers to such questions are of both theoretical and practical significance. Theoretically, it clarifies how universal turn-taking principles unfold within a diverse landscape of demographic and individual factors. Practically, these insights are important for the development of turn-taking models in conversational systems \cite{skantze2021}. While there have been a lot of efforts recently in developing predictive models of turn-taking based on human-human conversational data \cite{skantze2017towards, ekstedt2022, inoue2024}, the question remains how universal these models are and to what extent they should be conditioned on individual or demographic factors.  

\section{Materials and Method}

\subsection{Corpus}

The analyses were based on the Fisher corpus of spoken telephone conversations in US English \cite{cieri2004}, collected by the Linguistic Data Consortium (LDC). We used both part 1 and 2 of the dataset, containing 10,950 English telephone conversations with 11,684 different speakers. Each individual speaker was involved in up to four different conversations. For our analyses, we count each speaker in each conversation as a data point, resulting in 21,900 data points.

For the data collection, participants were randomly paired and engaged in spoken dialogues lasting up to 10 minutes, covering a wide range of topics that were given to the participants beforehand. One of the goals was to provide a representative distribution of subjects across a variety of demographic categories, such as sex, age, dialect region, and education.

Demographic information was taken directly from the labels provided in the CSV files.  In terms of speaker sex, females account for 56 \% of all calls.  The age distribution of participants is as follows: 22.5 \% are 16–24 years old, 29.4 \% are 25–34, 22.4 \% are 35–44, 15.8 \% are 45–54, and 9.5 \% are 55 or older.

\subsection{Turn-taking metric}

\begin{figure}[hbt!]
  \centering
  \includegraphics[width=\linewidth]{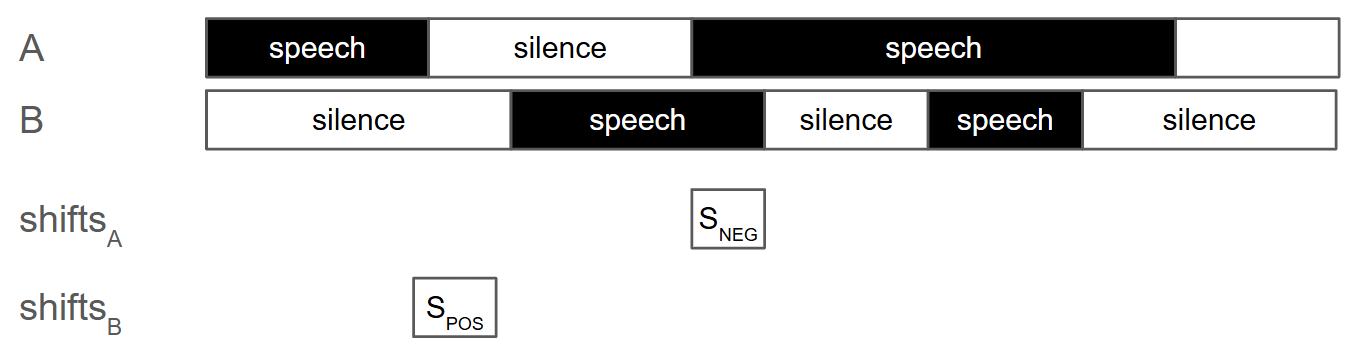}
  \caption{Illustration of how turn shifts were identified based on VAD data. $S_{pos}$ exemplifies a shift with positive TFO (a gap) attributed to speaker B. $S_{neg}$ exemplifies a shift with negative TFO (a between-speaker overlap) attributed to speaker A.}
  \label{fig:shift}
\end{figure}

In this study, we focus on the analysis of Transition Floor Offset (TFO) as a metric of turn-taking behavior. TFO refers to the temporal gap or overlap between the end of one speaker's turn and the beginning of the next speaker's turn. Negative TFO values indicate overlap, where the next speaker begins speaking before the current speaker has finished, whereas positive TFO values indicate a gap between turns. Zero TFO indicates a transition with no gap or overlap. To extract this metric, we first use a Voice Activity Detector (VAD) \cite{ekstedt2022} on each channel in the conversation. We then follow the procedure described in \cite{heldner2010} to identify turn shifts and their associated TFOs. We attribute each TFO to the next speaker taking the turn. A schematic illustration of this procedure is shown in Figure~\ref{fig:shift}.

All statistical analyses were conducted in R (version 4.4.1) \cite{R} using the \texttt{lme4} package \cite{lme}. Model parameters were estimated via restricted maximum likelihood (REML).

\section{Results}

\subsection{Overall distribution}

\begin{figure}[hbt!]
  \centering
  \includegraphics[width=\linewidth]{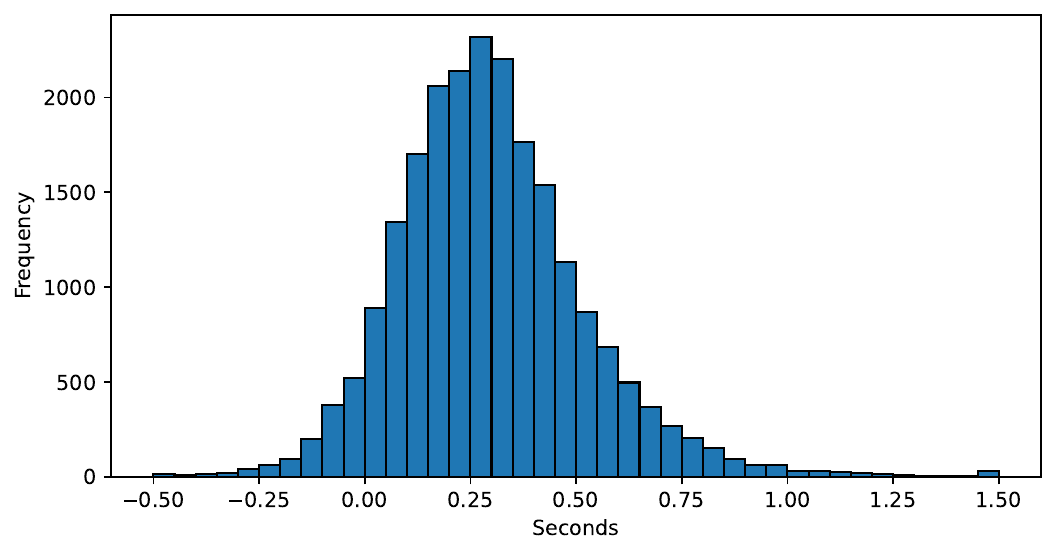}
  \caption{Histogram of average TFO per speaker and dialog.}
  \label{fig:tfo_dist}
\end{figure}

The average TFO for each speaker and dialogue was computed and the distribution is shown in Figure~\ref{fig:tfo_dist}. The average TFO is 0.300 s, which is similar that what has been reported in the literature \cite{stivers2009}. However, there is also a considerable variation between speakers and dialogs ($SD=0.228$). 

\subsection{Demographics: Effects of Sex, Age, and Education}

To investigate the effects of demographic factors on TFO, an ANOVA analysis was performed. The results are shown in Table~\ref{tab:anova_results}. All demographic factors show significant effects, but with small effect sizes: sex only explains about 3.2\% of the variation and age 1.3\%. The effect of education is negligible. 

\begin{table}[h]
\centering
\caption{ANOVA analysis of the effect of sex, age, and education on TFO with $\eta^2$ indicating effect sizes.}
\label{tab:anova_results}
\begin{tabular}{lrrrrr}
\toprule
{} &   sum\_sq &  F &  p &  $\eta^2$ \\
\midrule
Sex & 35.684 & 725.450 & $<0.001$ & 0.032 \\
Age & 14.434 & 293.442 & $<0.001$ & 0.013 \\
Education & 0.471 & 9.567 & 0.002 & 0.000 \\
Residual & 1070.735 & NaN & NaN & 0.955 \\
\bottomrule
\end{tabular}
\end{table}

\begin{figure}[hbt!]
  \centering
  \includegraphics[width=\linewidth]{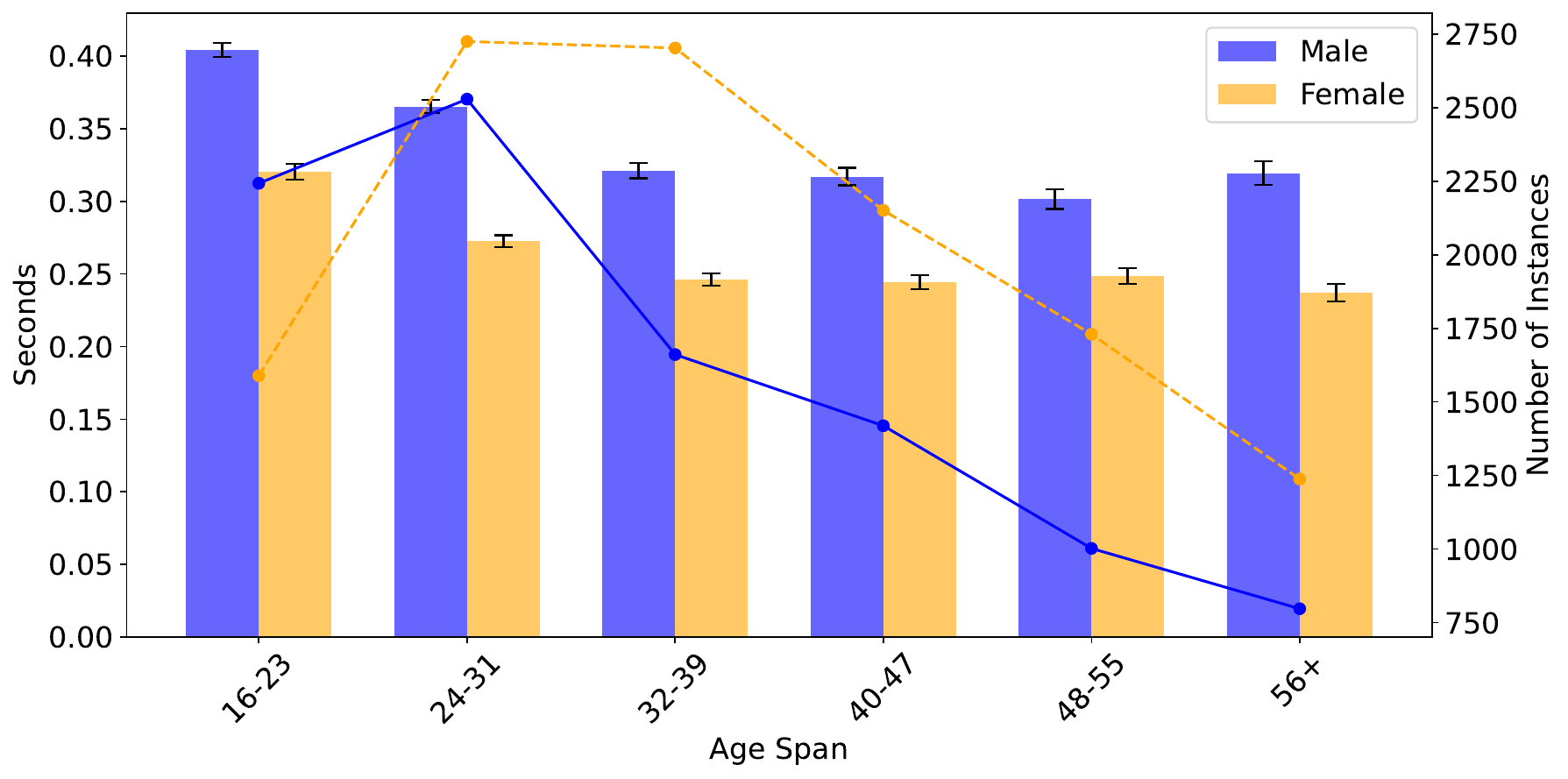}
  \caption{Bars: average TFO per speaker and dialog, split on age and sex, with SE bars. Lines: Number of instances.}
  \label{fig:tfo_age}
\end{figure}



The effect of both age and sex is shown in Figure~\ref{fig:tfo_age}. As can be seen, male speakers exhibited somewhat longer TFO. Older speakers tend to have somewhat shorter TFO, with a consistent sex effect. These two factors together account for roughly 4.5\% of the variation in TFO, while the residual (unexplained) portion is about 95.5\%.



\subsection{TFO and Topic}

\begin{figure}[hbt!]
  \centering
  \includegraphics[width=\linewidth]{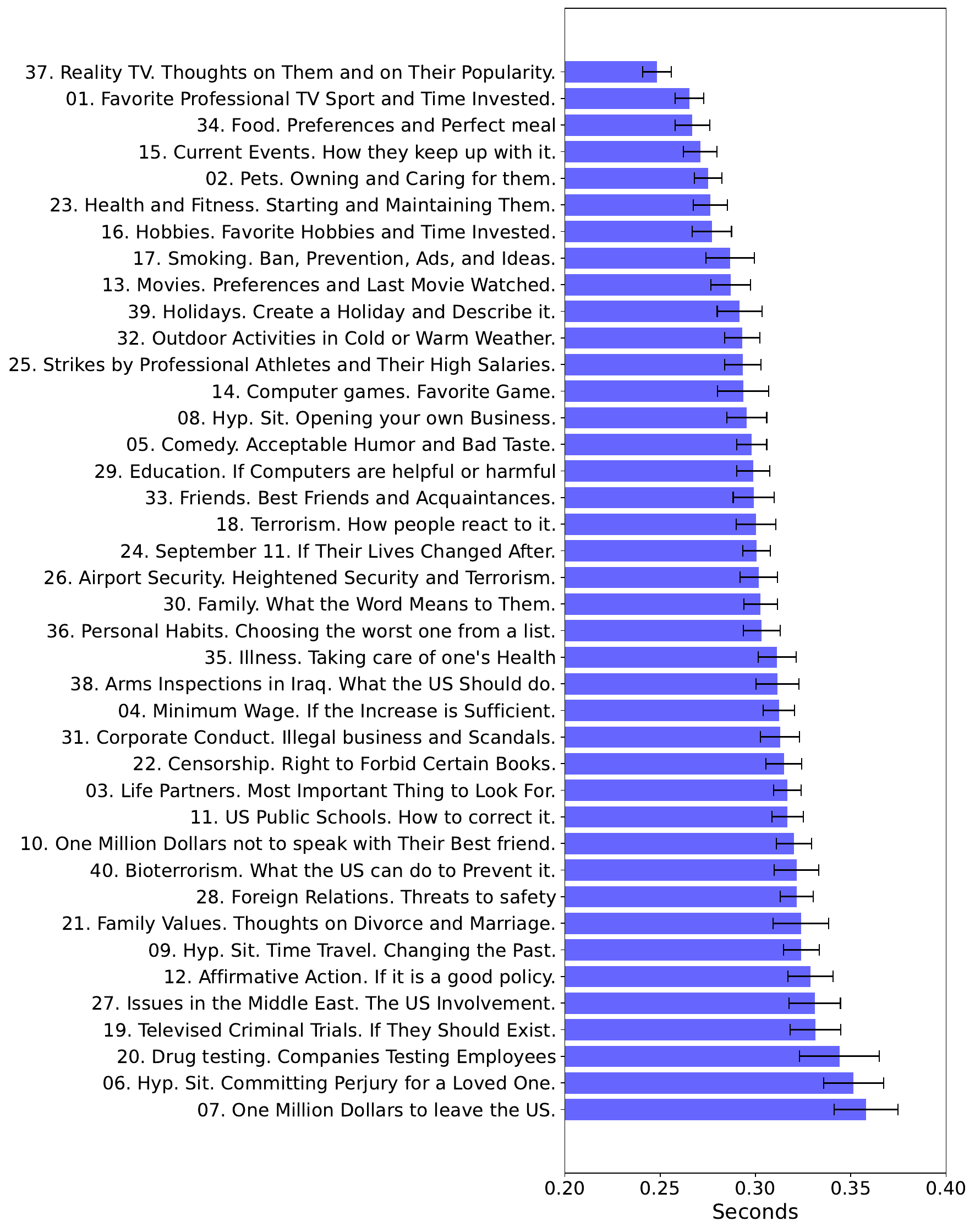}
  \caption{Average TFO per speaker and dialog, split on topic, with SE bars.}
  \label{fig:tfo_topic}
\end{figure}

Another factor that is likely to influence the turn-taking dynamics to a certain extent is the topic of the conversation. Since the speakers in the Fisher corpus data collection were asked to discuss specific topics, we can analyze this effect. The average TFO per topic is shown in Figure~\ref{fig:tfo_topic}, indicating that there is indeed an effect. Some specific topics yielded longer TFO while others displayed shorter values. In general, potentially sensitive topics, such as `hypothetical situations related to perjury', `drug testing', `life partners', `bioterrorism', resulted in longer TFO values, while lighter and more neutral topics, such as `current events', `health and fitness', `hobbies', `movies' yielded shorter TFO.

\subsection{Effects of Dyad and Speaker}

To examine the distinct contributions of individual speakers and dyadic interactions, we capitalized on the fact that many speakers in the Fisher dataset participated in multiple dyads. We fitted a linear mixed effect model, using `Age' (in years), `Sex' (male/female) and `Topic' (categorical with 40 levels) as fixed factors, and `Speaker ID' and `Dyad ID' as random factors. Note that each conversation has a unique `Dyad ID', but since each speaker in each conversation is treated as a data point, each `Dyad ID' will appear exactly twice (the same pair of speakers never had more than one conversation). Since Education only had a negligible effect in the analysis above, we left it out from this analysis.

\begin{table}[t]
\centering
\caption{Quantifying each factor’s contribution to TFO variance based on changes in $R^2$.}
\label{tab:r2}
\begin{tabular}{lr}
\toprule
Fixed effects &  Change in $R_m^2$ \\
\midrule
Sex & 0.0089 \\
Age   &  0.0081 \\
Topic & 0.0094 \\
\toprule
Random effects &  Change in $R_c^2$ \\
\midrule
Dyad & 0.415  \\
Speaker & 0.128 \\
\bottomrule
\end{tabular}
\end{table}

To be able to compare these various factors and estimate how much of the total variance each of them explains, we used Nakagawa and Schielzeth’s $R^2$ approach \cite{nakagawa2013} as implemented in the R package \textit{performance} \cite{Performance}. The marginal $R^2$ ($R_m^2$) represents the proportion of the total variance explained by the fixed effects only, whereas the conditional $R^2$ ($R_c^2$) is the proportion explained by the entire model (both fixed and random effects). 
To assess the unique contribution of each fixed effect, we compared a full model (which included all predictors) against a reduced model (omitting the predictor of interest). For each comparison, we calculated the change in $R_m^2$. This difference reflects how much additional variance that predictor explains beyond all other variables in the model.
Because multiple random effects were present, we separately examined how much variance each random effect accounted for. Specifically, we compared the full model to a model omitting the random intercept for `Dyad' or `Speaker' and calculated the change in $R_c^2$. These comparisons indicate how much variance is attributable to dyad-level differences vs. speaker-level differences. 

The result of this analysis is shown in Table~\ref{tab:r2}, revealing that ``dyadosyncrasy'' (`Dyad ID') is the most dominant factor. There is clearly also an effect of idiosyncrasy (`Speaker ID'), but it is considerably smaller. In other words, speakers in a specific dyad are much more similar to each other, in terms of TFO, than they are to themselves in a different dyad. Again, we see that the effects of the demographic factors are relatively small, even if they are statistically significant. 
Note that the effect of those factors are noticeably smaller than those reported in the ANOVA analysis in Table~\ref{tab:anova_results}. This is because ANOVA does not account for random effects or hierarchical structure, and calculates variance explained only within the fixed-effects framework. In contrast, the marginal $R^2$ from a mixed model reflects the proportion of total variance explained by fixed effects, including variance attributed to random effects and residuals in the denominator.



\section{Discussion}

In this study, we assessed the extent to which dyadic, individual and demographic factors can influence turn-taking behavior. The findings suggest that while the cognitive mechanisms supporting turn-taking may be universal, there is considerable variation in turn-taking behavior across speakers and dialogues, and different factors shape its precise implementation. Demographic factors appear to play some role. However, it is the ``dyadosyncratic'' and idiosyncratic factors that appear to have the most important effect, with the former having the most dominant effect.

In a broader sense, the ``dyadosyncratic'' variation observed may suggest pairs (rather than individuals) as the salient units of analysis in spontaneous dialogues. Paired speakers exhibit unique speech patterns that emerge from the interaction between the two individuals rather than from each person's independent characteristics. This result aligns with the notion of conversation as being best understood as a form of ``joint activity'' \cite{clark1996}. From such a perspective, dialogue does not result merely from the independent contributions of separate individuals but rather emerges through their mutual coordination and shared goals.

There is ample reason to believe that individual idiosyncrasies and dyadosyncratic interaction patterns are somewhat related. While individual idiosyncrasies define each speaker’s baseline tendencies, ``dyadosyncrasies'' arise from the synergy that unfolds when these patterns intersect. This higher-order phenomenon may reflect the evolving relationship, adaptation, and mutual influence between two speakers, resulting in a distinctive interactional signature that may not be fully captured by examining either participant alone.

As for the impact of demographic factors, such as sex, the results are consistent with what has been previously reported in the literature, with females exhibiting shorter TFO compared to males. Studies suggest that females tend to produce more overlaps than males, typically of a noncompetitive kind, serving to facilitate conversation by completing each other’s utterances and providing backchannel support. Consequently, women’s conversational style is often characterized as 'cooperative', while men's style is considered more ‘competitive’ \cite{coates1994}.

Regarding age, in previous studies this factor has been mainly assessed in relation to its developmental implications, e.g., \cite{ervin1979, casillas2016}. Here we were able to look at such a factor from another angle, considering that the minimum cut-off age in the dataset was 16 years. The results suggest that TFO tends to decrease for both Male and Female speakers during the lifespan. However, it remains uncertain whether TFO might rise again in later adulthood, as data become increasingly sparse at older ages, making any conclusions susceptible to sampling artifacts. An additional consideration is whether conversational norms have changed over time, raising the possibility that individuals from different generational cohorts may have developed distinct speaking styles.

TFO varied to some extent depending on the conversation topic. It is plausible that emotionally charged or complex topics may lead to longer turn transitions due to increased cognitive processing, politeness strategies, and social sensitivity. In complex topics, as this cognitive load increases, speakers may need longer turn transitions to formulate appropriate responses. For instance, it has been acknowledged that Wh-questions are slower than polar (yes–no) questions cross-linguistically, presumably because of the greater cognitive complexity of the responses involved \cite{levinson2015}. The relationship between the time of turn initiation and mutual common ground understanding has also been highlighted in previous studies. Shorter latencies often signal a greater degree of common ground understanding, that is, a mutual knowledge that is shared among interlocutors and that is known to be shared by them \cite{benus2011}.

For the development of predictive models of turn-taking in conversational systems \cite{skantze2021}, these results indicate that such models should take other factors into account than just the immediately preceding context. By conditioning the models on more long-term idiosyncratic and dyad-specific patterns, the model predictions are likely to improve. When applying these models to conversational systems, this can be challenging, as the system itself constitutes one part in the dyad on which the predictions should be conditioned.

Future research should examine a broader set of metrics such as overlaps, speech activity, and pause length. For instance, it has already been suggested that speech activity for the same speaker tends to vary considerably, depending on the pairing \cite{skantze2017}. The idiosyncratic nature of verbal backchannels has also been acknowledged, where some addressees are more active than others in providing feedback \cite{blomsma2024}. Although backchannels are distinguished from turn-taking cues in the literature, they can also be viewed as a form of cooperative overlap or, from a turn-taking perspective, as turn-yielding cues \cite{bertrand2007}. Utilizing a large, multimodal corpus of spontaneous interactions could further validate and extend the findings presented here \cite{reece2023}. It would be especially valuable to investigate how a broader set of demographic (e.g., race, social status, place of origin) and non-demographic factors (e.g., personality traits, conversational enjoyment) affect low-level turn-taking features.

Finally, some limitations should be acknowledged. Besides the fact that only English-speaking participants were involved, the dataset comprises conversations between individuals who had never encountered each other. This alone suggests that certain conversational aspects will be underrepresented. For instance, some topics may be easier to discuss with a familiar speaker than with a stranger. In addition, factors such as common ground \cite{clark1996} play a crucial role in interactions and should not be overlooked. It might very well be the case that the ``dyadsyncratic'' factor would play an even stronger role if the participants had known each other or were closely related \cite{cavalcanti2022}.

\section{Conclusion}

To our knowledge, this study is the first to systematically quantify how individual, demographic, and dyadic factors influence turn-taking behavior using a large-scale spoken dialogue dataset. Our findings reveal that while sex and age modestly shape the transition floor offset (TFO), dyadic interaction exerts the strongest influence, highlighting the need to treat dyads rather than individual speakers as the primary unit of analysis. Topic sensitivity also plays a role, with complex or emotionally charged topics slowing turn exchanges. These results reinforce the notion that conversation is a joint activity shaped by mutual adaptation rather than just a simple sum of isolated inter-speaker differences. 

\section{Acknowledgments}

This study was financed, in part, by the São Paulo Research Foundation (FAPESP), Brazil. Grant \#2024/06797-2. The research is connected to the thematic project titled “Multi-dimensional analysis of language, discourse, and society,” which is funded by FAPESP, grant \#22/05848-7. The work was also partially funded by Riksbankens Jubileumsfond (RJ) P20-0484, and Swedish Research Council (VR) 2020-03812.

\bibliographystyle{IEEEtran}
\bibliography{main}

\end{document}